\documentclass[11pt]{article}
\usepackage{fullpage, amsmath, amssymb}
\usepackage{graphicx}
\usepackage{cite}
\usepackage{hyperref}
\usepackage{color}
\usepackage{authblk}
\newcommand{\ket}[1]{|#1\rangle}
\newcommand{\bra}[1]{\langle #1|}
\newcommand{\Tr}{\mathrm{Tr}}
\newcommand{\D}{\mathrm{d}}
\newcommand{\diag}{\mathrm{diag}}
\begin{document}

\title{Quantum Coherence in a Quantum Heat Engine}
%Hai-Long Shi$^{4,5}$, Xiao-Hui Wang$^{1,7,*}$, Ming-Liang Hu$^{8}$, Si-Yuan Liu$^{2,6,7}$, Wen-Li Yang$^{1,6,7}$ and Heng Fan$^{2,3,9,*}$ }

\author[1,2,3]{Yun-Hao Shi}
\author[4,5]{Hai-Long Shi}
\author[1,7]{Xiao-Hui Wang \thanks{xhwang@nwu.edu.cn}}
\author[8]{Ming-Liang Hu}
\author[2,6,7]{Si-Yuan Liu}
\author[1,6,7]{Wen-Li Yang}
\author[2,3,9]{Heng Fan \thanks{hfan@iphy.ac.cn}}

\affil[1]{School of Physics, Northwest University, Xi'an 710127, China}
\affil[2]{Institute of Physics, Chinese Academy of Sciences, Beijing 100190, China}
\affil[3]{School of Physical Sciences $\&$ CAS Center for Excellence in Topological Quantum Computation, UCAS, Beijing 100190, China}
\affil[4]{State Key Laboratory of Magnetic Resonance and Atomic and Molecular Physics, Wuhan Institute of Physics and Mathematics, Chinese Academy of Sciences, Wuhan 430071, China}
\affil[5]{University of Chinese Academy of Sciences, Beijing 100049, China}
\affil[6]{Institute of Modern Physics, Northwest University, Xi'an 710127, China}
\affil[7]{Shaanxi Key Laboratory for Theoretical Physics Frontiers, Xi'an 710127, China}
\affil[8]{School of Science, Xi'an University of Posts and Telecommunications, Xi'an 710121, China}
\affil[9]{Songshan Lake Materials Laboratory, Dongguan 523808, China}

\maketitle

\begin{abstract}
We identify that quantum coherence is a valuable resource in the quantum heat engine, which is designed in
a quantum thermodynamic cycle assisted by a quantum Maxwell's demon. This demon is in a superposed state.
The quantum work and heat are redefined as the sum of coherent and incoherent parts in the energy representation.
The total quantum work and the corresponding efficiency of the heat engine can be enhanced due to the coherence consumption of
the demon. In addition, we discuss an universal information heat engine driven by quantum coherence. The extractable work of this heat engine is limited by the quantum coherence, even if it has no classical thermodynamic cost. This resource-driven viewpoint provides a direct and effective way to clarify the thermodynamic processes where the coherent superposition of states cannot be ignored.
\end{abstract}


\section{Introduction}
Quantum information theory~\cite{book-Nielsen2000,RMP-Vedral2002}
plays a crucial role in thermodynamics~\cite{NP-Sagawa2015,JPA-Goold2016,PRE-Goold2015,PRX-Strasberg2017,NC-Oppenheim2013,PNAS-Wehner2015,PRL-Oppenheim2015,PRA-Wehner2018,PRE-Goold2019,arXiv-Santos2017}.
Specifically, the rigorous frameworks for the quantification of quantum coherence~\cite{PRL-Plenio2014,RMP-Plenio2017,PR-Hu2018,PRL-Winter2016} and quantum correlations,
such as entanglement~\cite{PRL-Vedral1997} and discord~\cite{PRL-Zurek2001}, bring new insights into our understanding about quantum effects in thermodynamics.
Those new concepts created from quantum information motivate us to reconsider
the physical origin of thermodynamic notions such as work, heat, efficiency of heat engine etc.

On the other hand, remarkable progress of quantum thermodynamics began with the fundamental relations in non-equilibrium statistical mechanics known as the Jarzynski equality~\cite{PRL-Jarzynski1997,arXiv-Tasaki2000} or fluctuation theorems~\cite{RMP-Campisi2011,PRL-Potts2018,PRL-Sagawa2017,arXiv-Vedral2017}, which lead to the second law of thermodynamics as a corollary. The Maxwell's demon~\cite{Maxwell1871}, as an intermediary, bridges information and thermodynamics~\cite{RMP-Vedral2009}, such as Landauer's principle~\cite{Landauer1961,exp-Landauer2012} and generalized second laws~\cite{arXiv-Vedral2017,book-Sagawa2013,PRL-Quan2013,PRL-Mancino2018}. All of these support the view that information is a sort of physical existence, contributing to the adaptation of thermodynamics to quantum physics. Existing approaches to quantum thermodynamics are mainly based on operator formalisms~\cite{PRX-Strasberg2017,RMP-Campisi2011}, which provides a general framework for the thermodynamics of information. Another approach to quantum thermodynamics is path integral formalism~\cite{PRL-Quan2018} based on two-point measurement scheme~\cite{arXiv-Tasaki2000,arXiv-Kurchan2001}. This approach provides an effective way to calculate quantum work by utilizing various path integral techniques.
However, it is still difficult to figure out where information plays an important role and what kind of information is useful in a specific case,
since both classical information and quantum information facilitate thermodynamic processes.
It is interesting to explore such problems~\cite{NP-Sagawa2015,arXiv-Vedral2012}.

As is known, quantum coherence describes the departure of quantum mechanics from the classical realm.
It is a natural idea to interpret fundamental thermodynamic notions of quantum system by identifying the behaviour of quantum coherence and combining it with the same basic problem---Maxwell's demon~\cite{Maxwell1871}. Maxwell's demon has been a fundamental research question and topical issues for statistical mechanics and thermodynamics. Here we consider the object of quantum Maxwell's demon (QMD) in a quantum heat
engine.
This type of heat engine assisted by one quantum system has been extensively studied~\cite{RMP-Vedral2009,NJP-Campisi2017,E-Cherubim2019}. The other two types of engines are based on the parametric oscillator~\cite{PCCP-Salamon2009,IEEE-Stefanatos2017} and the invasive quantum measurement without feedback control~\cite{PRL-Campisi2019}, respectively. In this paper, we are more concerned about that when the demon is in a superposed state, how its quantum coherence disturbs the whole thermodynamic cycle of the heat engine. The ultimate goal is to explore the behaviour and role of quantum coherence in quantum thermodynamics.

This paper is organized as follows. Firstly, we discuss briefly the resource-theoretic framework of quantum coherence and introduce the QMD with coherence as the control unit in a quantum heat engine. Subsequently, the whole thermodynamic cycle of the heat engine assisted by QMD is studied, and the effects of quantum coherence on the fundamental thermodynamic notions are investigated. In addition, we reinvestigate the information heat engine (IHE) assisted by one memory with coherence, as an universal heat engine driven by quantum coherence. Meanwhile, we calculate the maximum work extractable with the coherence consumption. Finally, more general thermodynamics involving quantum coherence is considered.
%%%%%%%%%%%%%%%%%%%%%%%%%%%%%%%%%%%%%%%%5

\section{Quantum Coherence and Quantum Maxwell's Demon}
The rigorous framework for quantifying quantum coherence was first introduced by Baumgratz \emph{et al}. who proposed
four criteria for coherence measures~\cite{PRL-Plenio2014}. One well-defined measure is the relative entropy of coherence, which takes the form~\cite{PRL-Plenio2014,RMP-Plenio2017,PR-Hu2018,PRL-Winter2016}
\begin{equation}\label{eq:Cre}
C_r({\rho}):=\min_{\sigma\in \mathcal {I}}S(\rho\|\sigma)=S(\rho_{\mathrm{diag}})-S(\rho),
\end{equation}
where $\mathcal{I}$ denotes the set of incoherent states, $S(\rho)=-\Tr{(\rho\ln\rho)}$ is the von Neumann entropy~\cite{book-Nielsen2000} and $\rho_{\mathrm{diag}}$ is the state obtained from $\rho$ by deleting all the off diagonal elements. In the reference basis $\ket{i}$, $\rho_{\mathrm{diag}}$ can be expressed as $\sum_i\bra{i}\rho\ket{i}\ket{i}\bra{i}$. This entropic measure of coherence has a clear physical interpretation and many applications~\cite{PR-Hu2018}. For instance, quantum coherence can enhance the success probability in Grover search quantum algorithm~\cite{PRA-Shi2017}. In Ref.~\cite{PRA-Misra2016}, physical situations have been considered where the resource theories of coherence and thermodynamics play competing roles, especially the creation of quantum coherence comes at a cost of energy. Therefore, coherence can be viewed as a potential quantum resource in quantum physical processes. Exploring more applications of quantum coherence is of wide interests.

\begin{figure}[ht!]
	\centering
	\includegraphics[width=4.2in]{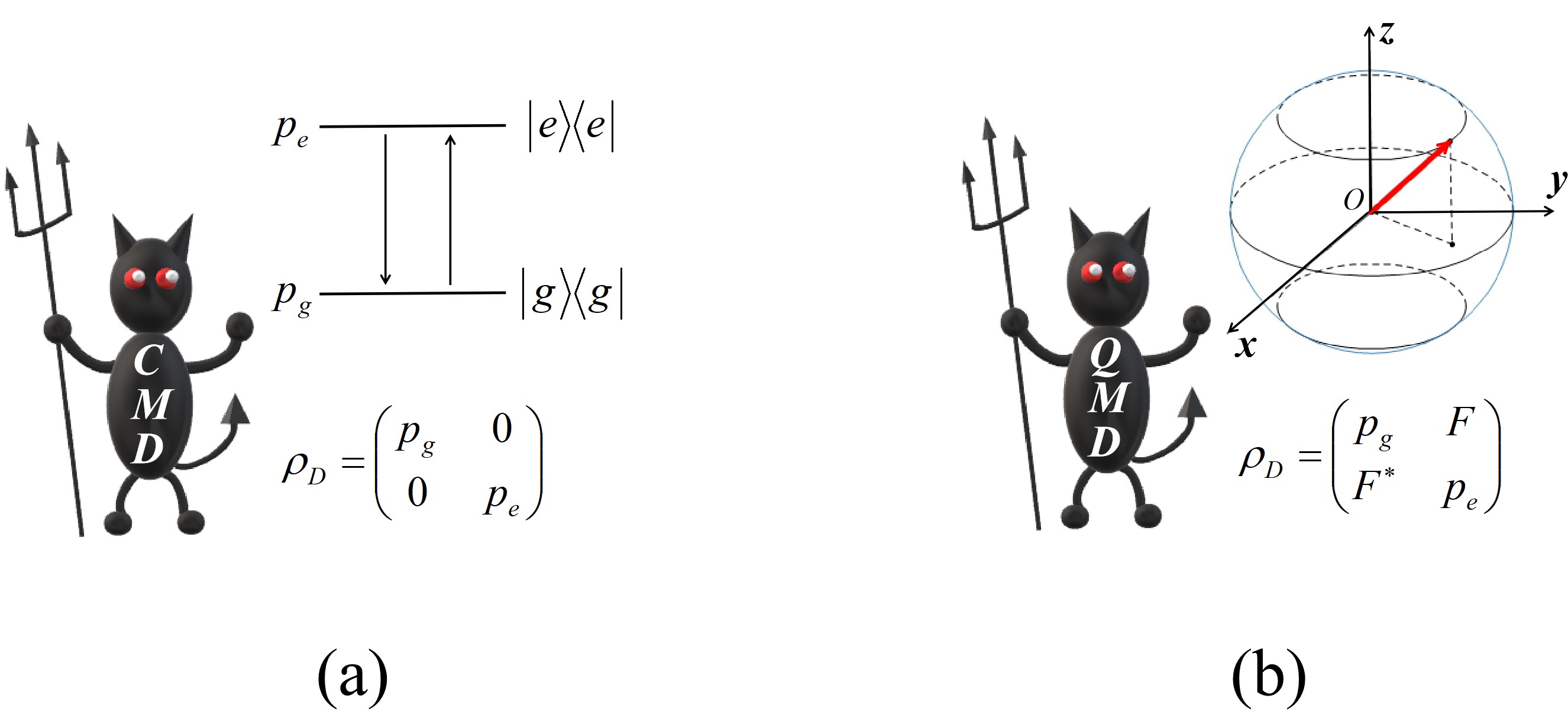}
	\caption{(color online) Classical Maxwell's demon (CMD) vs quantum Maxwell's demon (QMD). (a) CMD can be considered as a classical mixture of two definite eigenstates $\ket{g}$ and $\ket{e}$, while QMD with coherence should be preferred to a qubit described in the Bloch sphere, as shown in (b). Here, $p_g=(1+z)/2$, $p_e=(1-z)/2$, $F=(x-\mathrm{i}y)/2$ and $F^{\ast}=(x+\mathrm{i}y)/2$ are the classical probability distribution of ground state and excited state, and its off-diagonal elements, respectively. Note that $x^2+y^2+z^2\leq 1$, thus $|F|^2\leq p_gp_e$, where the equality is achieved if it is a pure quantum state.}
	\label{fig:CMDvsQMD}
\end{figure}

To investigate quantum coherence effect in quantum thermodynamics, we introduce a basic problem of statistical mechanics--the paradox of Maxwell's demon~\cite{Maxwell1871}, whose role of the information was firstly studied by Szilard~\cite{Szilard1929,RMP-Vedral2009}. A classical Maxwell's demon (CMD) can be described as an incoherent state, whereas a quantum Maxwell's demon (QMD) with off-diagonal elements is in a coherent state, namely
\begin{equation}\label{rhoD_i}
\rho_D=p_g\ket{g}\bra{g}+p_e\ket{e}\bra{e}+F\ket{g}\bra{e}+F^{\ast}\ket{e}\bra{g},
\end{equation}
where $p_g$ and $p_e=1-p_g$ are the probability distributions in the ground state $\ket{g}$ and excited state $\ket{e}$, respectively. QMD can be viewed as an arbitrary single-qubit system expressed in the Bloch sphere (see Fig.~\ref{fig:CMDvsQMD}). The off-diagonal elements $F$ and $F^{\ast}$ are introduced to embody the quantum coherence. The QMD and its related second law-like equalities (inequalities) have already been experimentally verified by using various systems in recent years~\cite{exp-MD-Camati2016,exp-MD-Vedral2016,exp-MD-Murch2018,exp-MD-Elouard2017,exp-MD-Koski2015,exp-MD-Koski2014,exp-MD-Sagawa2010}.

According to Eqs.~\eqref{eq:Cre} and~\eqref{rhoD_i}, the coherence of QMD can be written as
\begin{equation}\label{CreQMD}
C_r(\rho_D)=H(p_g)-H(\lambda_{+}),
\end{equation}
where $H(x)=H(1-x)=-x\ln{x}-(1-x)\ln{(1-x)}$ is the Shannon entropy function~\cite{book-Nielsen2000} and $\lambda_{\pm}=\frac{1}{2}(1\pm\sqrt{(p_g-p_e)^2+4|F|^2})$ are the eigenvalues of $\rho_D$. If we set $\rho_D$ as the initial state of QMD in the heat engine $\rho_D^i$, the initial coherence is exactly in the form of Eq.~\eqref{CreQMD}.

In what follows, we will focus on the analysis of the whole quantum thermodynamic cycle assisted by QMD based on quantum Szilard engine (QSE)~\cite{PRL-Kim2011,PRE-Dong2011} and reveal the behavior and role of quantum coherence.
%%%%%%%%%%%%%%%%%%%%%%%%%%%%%%%%%%%%%%%%

\section{Quantum Maxwell's Demon with Coherence Assisted Quantum Thermodynamic Cycle}
As a carrier of coherence, QMD is typically designed to drive a QSE~\cite{PRL-Kim2011,Comment-QSE}, which consists of a closed box with definite size $L$ and one system $S$ (such as a single molecule). The demon in our thermodynamic cycle is in contact with a lower temperature heat bath while the system's bath is at a higher temperature. Although the system is in equilibrium, the QMD is in a nonequilibrium state, as a unit to control expansion. In the cycle, the quantum coherence of the demon is consumed continuously, contributing to the extra work.

\begin{figure*}[ht]
	\centering
	\includegraphics[width=15.5cm]{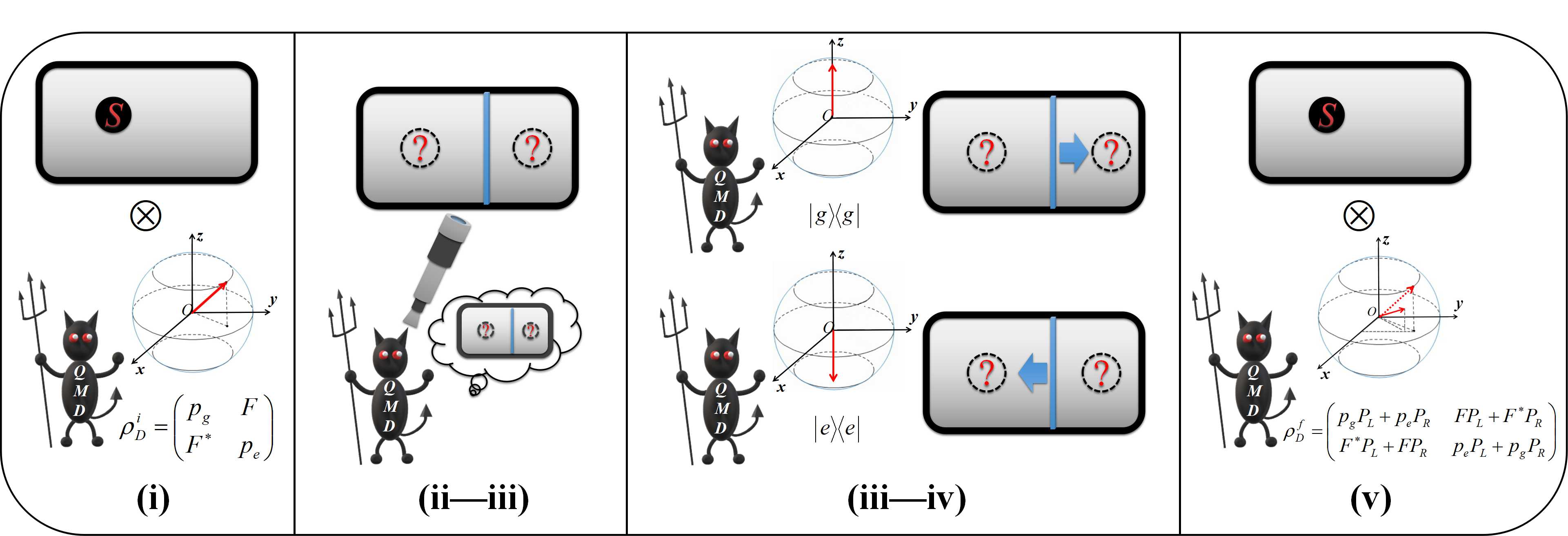}
	\caption{Schematic diagram of the quantum thermodynamic cycle assisted by quantum Maxwell's demon (QMD). (i) A system $S$ is prepared in a closed box with size $L$ while QMD with coherence is isolated from the box. (ii---iii) A wall, depicted by a vertical blue bar, is isothermally inserted at location $l$. And the measurement is performed later to register the information of the system $S$ into QMD's memory. (iii---iv) According to the state of the quantum memory, the wall controlled by QMD undergoes a reversible and isothermal expansion until it reaches equilibrium. (v) To complete this thermodynamic cycle, the wall is removed isothermally. At the moment, the state of QMD is overwhelmingly different from the initial state and its coherence has been consumed. }
	\label{fig:sch}
\end{figure*}

The whole thermodynamic cycle is briefly shown in Fig.~\ref{fig:sch}, which is split into five stages: (i) Initial state, (ii) Insertion, (iii) Measurement, (iv) Expansion, and (v) Removal. At the stage of insertion, a wall is isothermally inserted at location $l$. We denote $P_L$ as the quantum probability of the system $S$ on the left after insertion , then $P_R=1-P_L$ is the probability of $S$ on the right. What the QMD should do are performing a global measurement both on the system $S$ and itself (the controlled-NOT operation~\cite{PRA-Zurek2003,PRL-Quan2006,PRE-Dong2011}), and then controlling the expansion of $S$. When the wall is removed and this thermodynamic cycle is completed, $S$ returns to its initial state of equilibrium. In the presence of quantum coherence, it is insufficient to know its diagonal part. Instead, we describe the evolution of the whole system in terms of the full density matrix.

Stage (\romannumeral1): Initial state.---
Initially, the system $S$, prepared in a closed box with size $L$, contacts with the reservoir at temperature $T=(k_{\text{B}}\beta)^{-1}$ so that it is in equilibrium. With the initial Hamiltonian $H_S=\sum_{n}E_n(L)\ket{\psi_n(L)}\bra{\psi_n(L)}$, the density matrix of the initial state of the total system can be written as a product state
\begin{alignat}{2}\label{rho_i}
\begin{split}
\rho^{i}&=\rho^{i}_S\otimes\rho^{i}_D\\
&=\sum_{n}\frac{e^{-\beta E_n(L)}}{Z(L)}\ket{\psi_n(L)}\bra{\psi_n(L)}\otimes\rho^{i}_D,
\end{split}
\end{alignat}
where $Z(L)=\sum_{n}e^{-\beta E_n(L)}$, $E_n(L)$ and $\ket{\psi_n(L)}$ are the partition function, $n$-th eigenenergy and eigenstate of the system $S$ in this box with size $L$.
Here, the initial density matrix of QMD, namely $\rho_{D}^{i}$ takes the form of Eq.~\eqref{rhoD_i}.

The system $S$ will always be in the isothermal process with the reservoir, leading to its thermalization so that all the coherence among energy levels vanishes~\cite{PRL-Kim2011}. Even so, it is necessary to describe the dynamics of the total system in terms of the full density matrix, because our QMD with off-diagonal elements is still coherent at another temperature $T_D=(k_{\text{B}}{\beta_D})^{-1}$.

Here we emphasize that our discussion doesn't depend on a specific model. For instance, the box doesn't have to be an infinite potential well model, i.e., the eigenenergies and eigenfunctions of the system $S$ are $E_n(L)=\hbar^2\pi^2n^2/(2mL^2)$ and $\langle x|\psi_n(L)\rangle=\sqrt{2/L}\sin(\pi nx/L)$, where the quantum number $n$ ranges from $1$ to $\infty$.

Stage (\romannumeral2): Insertion.---
After preparing initial state, a wall is then isothermally inserted at a certain position $l$. In the classical situation, the position of the system $S$ is definite, so either it is on the left side or stay on the right side, while in the quantum case the system is simultaneously on both sides of the wall, until its location is determined by the measurement of the observer. Therefore, when the wall insertion is completed, the total density matrix can be expressed as a mixture, namely
\begin{equation}\label{rho_ins}
\rho^{\text{ins}}=[P_L\rho^L(l)+P_R\rho^R(L-l)]\otimes\rho^{i}_D,
\end{equation}
where $\rho^L(l)=\sum_{n}\frac{e^{-\beta E_n(l)}}{Z(l)}\ket{\psi_n^L(l)}\bra{\psi_n^L(l)}$ and $\rho^R(L-l)=\sum_{n}\frac{e^{-\beta E_n(L-l)}}{Z(L-l)}\ket{\psi_n^R(L-l)}\bra{\psi_n^R(L-l)}$ are the independent densities of $S$ on the left and right sides, $P_L=Z(l)/[Z(l)+Z(L-l)]$ and $P_R=1-P_L$ are the corresponding quantum probabilities, which differ from the classical probabilities $P_L^{\text{classical}}=l/L$ and $P_R^{\text{classical}}=(L-l)/L$~\cite{PRE-Dong2011}. In this stage, QMD is still out of work, which is the same as CMD. And because the system $S$ has no coherence, thus the results (work, heat, etc.) in this stage only relate with diagonal part of $S$, as in the case of no coherence in Ref.~\cite{PRL-Kim2011}.

Stage (\romannumeral3): Measurement.---
Now the demon will perform a global measurement, which can be described by the controlled-NOT operation~\cite{PRL-Quan2006,PRE-Dong2011}, namely
\begin{alignat}{2}\label{C-NOT}
\begin{split}
U&=\sum_n[\ket{\psi_n^L(l)}\bra{\psi_n^L(l)}\otimes(\ket{g}\bra{g}+\ket{e}\bra{e})\\
&\quad+\ket{\psi_n^R(L-l)}\bra{\psi_n^R(L-l)}\otimes(\ket{e}\bra{g}+\ket{g}\bra{e})].
\end{split}
\end{alignat}
It is easy to verify that $U$ is a unitary operator, which is different from the positive operator valued measure (POVM) in previous discussions~\cite{PRL-Park2013,PRL-Sagawa2008,PRL-Sagawa2009}. Here, the physical process corresponding to $U$ is that the demon will keep the original state if the system $S$ with $n$ levels is on the left side, while the demon flips its state when $S$ is on the right. After this operation, the total density matrix is given by

\begin{alignat}{2}\label{rho_mea}
\begin{split}
\rho^{\text{mea}}&=U\rho^{\text{ins}}U^{\dagger}\\
&=P_L\rho^L(l)\otimes(p_g\ket{g}\bra{g}+p_e\ket{e}\bra{e}+F\ket{g}\bra{e}+F^{\ast}\ket{e}\bra{g})\\
&\quad\!+\!P_R\rho^R(L-l)\otimes(p_e\ket{g}\bra{g}+p_g\ket{e}\bra{e}+F^{\ast}\ket{g}\bra{e}+F\ket{e}\bra{g})\\
&=[p_gP_L\rho^L(l)\!\!+\!\!p_eP_R\rho^R(L\!-\!l)]\!\otimes\!\ket{g}\bra{g}\!\!+\!\![p_eP_L\rho^L(l)\!\!+\!\!p_gP_R\rho^R(L\!-\!l)]\!\otimes\!\ket{e}\bra{e}\\
&\quad\!\!+\!\![F\!P_L\rho^L(l)\!\!+\!\!F^{\ast}\!P_R\rho^R(L\!-\!l)]\!\otimes\!\ket{g}\bra{e}\!\!+\!\![F^{\ast}\!P_L\rho^L(l)\!\!+\!\!F\!P_R\rho^R(L\!-\!l)]\!\otimes\!\ket{e}\bra{g},
\end{split}
\end{alignat}
which means QMD is now correlated with the system $S$, contributing to the work to the outside.

Compared with CMD, QMD seem to be ``blurry-eyed'', so that it can't exactly distinguish which side the system $S$ stays on. In the limits of $F\rightarrow0$ and $T_D\rightarrow0$, $\rho^{\text{mea}}\rightarrow P_L\rho^L(l)\otimes\ket{g}\bra{g}+P_R\rho^R(L-l)\otimes\ket{e}\bra{e}$, which implies that QMD is ``cured'', i.e., its ground state $\ket{g}$ corresponds to the system $S$ on the left and vice versa, as the case of CMD.

For registering the information of the system $S$ into QMD's memory, the outside agent have to do work on the total system in this process, leading to the change of the total internal energy. The mean value of this work, namely $\langle W_{\text{mea}}\rangle=\Tr[(H_S\!+\!H_D)(\rho^{\text{mea}}\!-\!\rho^{\text{ins}})]=P_R(p_g-p_e)\Delta$, has nothing to do with the off-diagonal elements, because the flip of off-diagonal part makes no contribution to the change of internal energy. However, the off-diagonal part has affected the total density matrix, giving rise to a more complex form about entropy and heat as the following discussions. Here, $H_S$ and $H_D$ are the Hamiltonian of the system and the demon, respectively. We assume that the demon is a two-level system so that $H_D=E_g\ket{g}\bra{g}+E_e\ket{e}\bra{e}$ and $\Delta=E_e-E_g$.

Stage (\romannumeral4): Expansion.---
In this stage, the expansion of the system $S$ is slowly enough to ensure the process reversible and isothermal, which is controlled by QMD according to its memory. The wall will be allowed to move to the right side at a final position $l_g$ if QMD is in the ground state and move to the right side at a final position $l_e$ if the demon is in the excited state (see intuitively in Fig.~\ref{fig:sch} (iii---iv)). In accordance with the above analysis, the evolution operator of the controlled expansion can be mathematically constructed as
\begin{alignat}{2}\label{O_exp}
\begin{split}
O_{\text{exp}}&\!=\!\!\sum_n\!\textstyle\Bigg\{\!\bigg[\!\sqrt{\!\frac{P_n(l_g)}{P_n(l)}}\ket{\psi_n^L(l_g)}\bra{\psi_n^{L}(l)}\!+\!\!\sqrt{\!\frac{P_n(L-l_g)}{P_n(L-l)}}\ket{\psi_n^R(L\!-\!l_g)}\bra{\psi_n^R(L\!-\!l)}\bigg]\!\!\otimes\!\ket{g}\bra{g}\\
&\quad+\!\textstyle\bigg[\!\sqrt{\!\frac{P_n(l_e)}{P_n(l)}}\ket{\psi_n^L(l_e)}\bra{\psi_n^L(l)}\!+\!\sqrt{\!\frac{P_n(L-l_e)}{P_n(L-l)}}\ket{\psi_n^R(L\!-\!l_e)}\bra{\psi_n^R(L\!-\!l)}\bigg]\!\!\otimes\!\ket{e}\bra{e}\!\Bigg\},
\end{split}
\end{alignat}
where $P_n(x)=\frac{e^{-\beta E_n(x)}}{Z(x)}$. We thus obtain the total density matrix after expansion:
\begin{alignat}{2}\label{rho_exp}
\begin{split}
\rho^{\text{exp}}&\!=O_{\text{exp}}\rho^{\text{mea}}O_{\text{exp}}^{\dagger}\\
&=\!\big[p_gP_L\rho^L(l_g\!)\!\!+\!p_eP_R\rho^R(L\!\!-\!\!l_g\!)\big]\!\otimes\!\ket{g}\bra{g}\!\!+\!\!\big[p_eP_L\rho^L(l_e\!)\!\!+\!p_gP_R\rho^R(L\!\!-\!\!l_e)\big]\!\otimes\!\ket{e}\bra{e}\\
&\quad+\big[FP_L\rho^L_{l_g,l_e}+F^{\ast}P_R\rho^R_{l_g,l_e}\big]\otimes\ket{g}\bra{e}+\big[F^{\ast}P_L\rho^L_{l_e,l_g}+FP_R\rho^R_{l_e,l_g}\big]\otimes\ket{e}\bra{g},
\end{split}
\end{alignat}
where $\rho^L(l_g)\!=\!\sum_n\!P_n(l_g)\ket{\psi_n^L(l_g)}\bra{\psi_n^{L}(l_g)}$ and $\rho^R(l_e)\!=\!\sum_n\scriptstyle\!P_n(l_e)\ket{\psi_n^R(l_e)}\bra{\psi_n^{R}(l_e)}$ are the post-expansion densities controlled by the diagonal parts $\ket{g}\bra{g}$ and $\ket{e}\bra{e}$, respectively. Similarly, the items $\rho^L_{l_g,l_e}\!=\!\sum_n\scriptstyle\!\sqrt{P_n(l_g)P_n(l_e)}\ket{\psi_n^L(l_g)}\bra{\psi_n^{L}(l_e)}$ and $\rho^R_{l_g,l_e}\!=\!\sum_n\scriptstyle\!\sqrt{P_n(L\!-\!l_g)P_n(L\!-\!l_e)}\ket{\psi_n^{R}(L\!-\!l_g)}\bra{\psi_n^{R}(L\!-\!l_e)}$ in Eq.~\eqref{rho_exp} can be viewed as the post-expansion densities controlled by the off-diagonal part $\ket{g}\bra{e}$, whereas $\rho^L_{l_e,l_g}\!=\!\sum_n\scriptstyle\sqrt{P_n(l_e)P_n(l_g)}\ket{\psi_n^L(l_e)}\bra{\psi_n^{L}(l_g)}$ and $\rho^R_{l_e,l_g}\!=\!\sum_n\scriptstyle\sqrt{P_n(L\!-\!l_e)P_n(L\!-\!l_g)}\ket{\psi_n^{R}(L\!-\!l_e)}\bra{\psi_n^{R}(L\!-\!l_g)}$ are controlled by $\ket{e}\bra{g}$.

In classical physics, the system must be on one side of the wall with another side empty after insertion, thus the wall is doomed to be moved to an end boundary of the box due to expansion. In contrast, quantum system can be simultaneously on both sides of the wall, thus the equilibrium position is somewhere in the box rather than the boundary. The condition of equilibrium, in mechanics, is that the wall has equal and opposite forces on the two sides, i.e. $F^L_{\text{eq}}=F^R_{\text{eq}}$.
%The comparison between Eq.~\eqref{rho_mea} and Eq.~\eqref{rho_exp} in order indicates that the diagonal parts physically contribute to the transitions

Stage (\romannumeral5): Removal.---
For purpose of the thermodynamic cycle, the system $S$ must be reset to its own initial state, i.e. $\rho^{i}_S=\sum_{n}\frac{e^{-\beta E_n(L)}}{Z(L)}\ket{\psi_n(L)}\bra{\psi_n(L)}$. The corresponding physical process is the wall inserted on the stage (\romannumeral2) will be removed and all the ensembles of the system $S$ will evolve into $\rho^{i}_S$, namely
\begin{equation}\label{rho_rem}
\rho^{\text{rem}}=\sum_{n}\frac{e^{-\beta E_n(L)}}{Z(L)}\ket{\psi_n(L)}\bra{\psi_n(L)}\otimes\rho_D^{f}.
\end{equation}
where
\begin{alignat}{2}\label{rhoD_rem}
\begin{split}
\rho_D^{f}&=\big[(p_gP_L+p_eP_R)\ket{g}\bra{g}+(p_eP_L+p_gP_R)\ket{e}\bra{e}\\
&\quad+(FP_L+F^{\ast}P_R)\ket{g}\bra{e}+(F^{\ast}P_L+FP_R)\ket{e}\bra{g}\big].
\end{split}
\end{alignat}
In analogy with Eq.~\eqref{O_exp}, we can also construct an operation $O_{\text{rem}}$ to obtain Eq.~\eqref{rho_rem}, i.e., $\rho^{\text{rem}}=O_{\text{rem}}\rho^{\text{exp}}O_{\text{rem}}^{\dagger}$ with

\begin{alignat}{2}\label{O_rem}
\begin{split}
O_{\text{rem}}&\!=\!\sum_n\!\textstyle\Bigg\{\bigg[\!\sqrt{\!\frac{P_n(L)}{P_n(l_g)}}\ket{\psi_n(L)}\bra{\psi_n^L(l_g)}\!+\!\sqrt{\!\frac{P_n(L)}{P_n(L-l_g)}}\ket{\psi_n(L)}\bra{\psi_n^R(L\!-\!l_g)}\bigg]\!\!\otimes\!\ket{g}\bra{g}\\
&\quad+\!\textstyle\bigg[\!\sqrt{\!\frac{P_n(L)}{P_n(l_e)}}\ket{\psi_n(L)}\bra{\psi_n^L(l_e)}\!+\!\sqrt{\!\frac{P_n(L)}{P_n(L-l_e)}}\ket{\psi_n(L)}\bra{\psi_n^R(L-l_e)}\bigg]\!\!\otimes\!\ket{e}\bra{e}\Bigg\}.
\end{split}
\end{alignat}
Here we emphasize that the system $S$ has been reset to thermal equilibrium state with no coherence, but this is not the requirement for QMD. The state of QMD remains coherent, but obviously it's much less coherent than it was at the beginning. This implies that the coherence consumption may be a direct linkage to work and heat in the whole thermodynamic cycle, as discussed below.
%%%%%%%%%%%%%%%%%%%%%%%%%%%%%%%%%%%%%%%

\section{Enhancing Efficiency via Coherence Consumption of Quantum Maxwell's Demon}
What we have talked above is the evolution of state of the total system. Because of the coherence, it is insufficient to know its diagonal part. Instead, we describe the evolution of our system in terms of the full density matrix, which is split into five, namely Stage (\romannumeral1-\romannumeral5). As seen in the follows, this approach is in support of studying what we are more concerned, i.e., whether the efficiency of this quantum thermodynamics cycle could be enhanced due to the coherent QMD.

First of all, let us calculate the change of total internal energy $\langle\Delta E_{\text{tot}}\rangle$ by taking $\rho^{\text{rem}}$ as the final density $\rho^{f}$ at the end of this thermodynamics cycle. Focusing on the initial and final states, we can obtain
\begin{equation}\label{DeltaU_tot}
\langle\Delta E_{\text{tot}}\rangle=\langle E_{f}\rangle-\langle E_{i}\rangle=\Tr\big[(\rho^{f}-\rho^{i})(H_S+H_D)\big]=P_R(p_g-p_e)\Delta,
\end{equation}
where $H_D=E_g\ket{g}\bra{g}+E_e\ket{e}\bra{e}$ is the demon's Hamiltonian and $\Delta=E_e-E_g$ is the gap of the two levels of QMD. Here, the result is independent of $H_S$, namely the form of the Hamiltonian of the system $S$ is inessential during the cycle.  Note that $\langle\Delta E_{\text{tot}}\rangle=\langle W_{\text{mea}}\rangle$, i.e., the change of total internal energy merely results from the work done by the outside agent during measurement, as what mentioned before. Eq.~\eqref{DeltaU_tot} also confirms the fact that the change of internal energy does not involve the off-diagonal part at all.

Another important thermodynamic quantity is the total heat $\langle Q_{\text{tot}}\rangle$ absorbed from the outside, which is associated with the total entropy change due to the reversibility, namely
\begin{alignat}{2}\label{Q_tot}
\begin{split}
\langle Q_{\text{tot}}\rangle&\!=\! k_{\text{B}}T\big[S(\rho^{f})-S(\rho^{i})\big]\\
&\!=\! k_{\text{B}}T\!\bigg[S\textstyle\Big(\!\sum_{n}\!\frac{e^{-\beta E_n(L)}}{Z(L)}\ket{\psi_n(L)}\bra{\psi_n(L)}\!\otimes\!\rho_D^{f}\!\Big)\!-\!S\Big(\!\sum_{n}\!\frac{e^{-\beta E_n(L)}}{Z(L)}\ket{\psi_n(L)}\bra{\psi_n(L)}\!\otimes\!\rho_D^{i}\!\Big)\!\bigg]\\
&\!=\! k_{\text{B}}T\big[S(\rho_D^{f})-S(\rho_D^{i})\big]\\
&\!=\! T(\Delta S_{\text{c}}+k_{\text{B}}\Delta C_r),
\end{split}
\end{alignat}
where the second to the last line follows from the subadditivity equality for von Neumann entropy~\cite{book-Nielsen2000} and the last line follows from the definition of the coherence consumption $\Delta C_r:=C_r(\rho^i)-C_r(\rho^f)=C_r(\rho^i_D)-C_r(\rho^f_D)$, and $\Delta S_{\text{c}}:=k_{\text{B}}\Delta S_{\diag}=k_{\text{B}}[S(\rho^f_{\diag})-S(\rho^i_{\diag})]=k_{\text{B}}[S({\rho^f_D}_{\text{diag}})-S({\rho^i_D}_{\diag})]$. According to what we discussed above, the initial density matrix and the final density matrix are expressed as
\begin{equation}\label{matrixi}
\rho^i_D=\left(
\begin{array}{cc}
p_g & F \\
F^* & p_e \\
\end{array}
\right)
\end{equation}
and
\begin{equation}\label{matrixf}
\rho^f_D=\left(
\begin{array}{cc}
p_gP_L+p_eP_R & FP_L+F^*P_R \\
F^*P_L+FP_R & p_gP_R+p_eP_R \\
\end{array}
\right),
\end{equation}
respectively. Hence, one can obtain the change of classical entropy
\begin{alignat}{2}
\begin{split}
\Delta S_{\text{c}}&=\!k_{\text{B}}\big[S({\rho^f_D}_{\text{diag}})-S({\rho^i_D}_{\diag})\big]\\
&=\!k_{\text{B}}\big[p_g\!\ln{p_g}\!+\!p_e\!\ln{p_e}\!-\!{(p_gP_L\!\!+\!\!p_eP_R)}\ln{(p_gP_L\!\!+\!\!p_eP_R)}\!-\!{(p_eP_L\!\!+\!\!p_gP_R)}\ln{(p_eP_L\!\!+\!\!p_gP_R)}\big]
\end{split}
\end{alignat}
and the coherence consumption
\begin{equation}\label{coherence consumption}
\Delta C_r=C_r(\rho^i_D)-C_r(\rho^f_D)=S({\rho^f_D})-S({\rho^i_D})-\Delta S_{\text{c}}/k_{\text{B}}.
\end{equation}
Note that in the classical process, the heat without coherence can be denoted as $\langle Q_{\text{incoh}}\rangle=T\Delta S_{\diag}$, which just depends on the diagonal part. In general, we specify
\begin{equation}\label{Q_tot=incoh+coh}
\langle Q_{\text{tot}}\rangle=\langle Q_{\text{incoh}}\rangle+\langle Q_{\text{coh}}\rangle
\end{equation}
as the total heat absorbed from the outside, where the coherent item $\langle Q_{\text{coh}}\rangle=k_{\text{B}}T\Delta C_r$ is proportional to the coherence consumption $\Delta C_r$.

Then if we still believe that the first law holds, i.e., $\langle\Delta E_{\text{tot}}\rangle=\langle Q_{\text{tot}}\rangle-\langle W_{\text{tot}}\rangle$, the total work $\langle W_{\text{tot}}\rangle$ done by the system to the outside will be expressed as
\begin{equation}\label{W_tot}
\langle W_{\text{tot}}\rangle=T(\Delta S_{\text{c}}+k_{\text{B}}\Delta C_r)-P_R(p_g-p_e)\Delta.
\end{equation}
An alternative derivation of this result can be constructed by using $\langle W_{\text{tot}}\rangle=\Delta F$ with the standard free energy $F(\rho)=\Tr{(H\rho)}-TS(\rho)$~\cite{NP-Sagawa2015,PRL-Oppenheim2015}, which is a special case of the general free energy $F_\alpha(\rho)=F_{\text{eq}}+TD_\alpha(\rho\|\rho_{\text{eq}})$ based on quantum Renyi entropies $D_\alpha(\rho\|\rho_{\text{eq}})$ when $\alpha\rightarrow1$~\cite{PRA-Wehner2018,PRE-Goold2019,arXiv-Santos2017}.

In analogy to Eq.~\eqref{Q_tot=incoh+coh}, the total work can be split into
\begin{equation}\label{W_tot=incoh+coh}
\langle W_{\text{tot}}\rangle=\langle W_{\text{incoh}}\rangle+\langle W_{\text{coh}}\rangle,
\end{equation}
where the incoherent work is $\langle W_{\text{incoh}}\rangle=\langle Q_{\text{incoh}}\rangle+\langle\Delta E_{\text{tot}}\rangle=T\Delta S_{\text{c}}-P_R(p_g-p_e)\Delta$ and $\langle W_{\text{coh}}\rangle=\langle Q_{\text{coh}}\rangle=k_{\text{B}}T\Delta C_r$ represents the coherent part.

\begin{figure}[ht]
	\centering
	\includegraphics[width=3.5in]{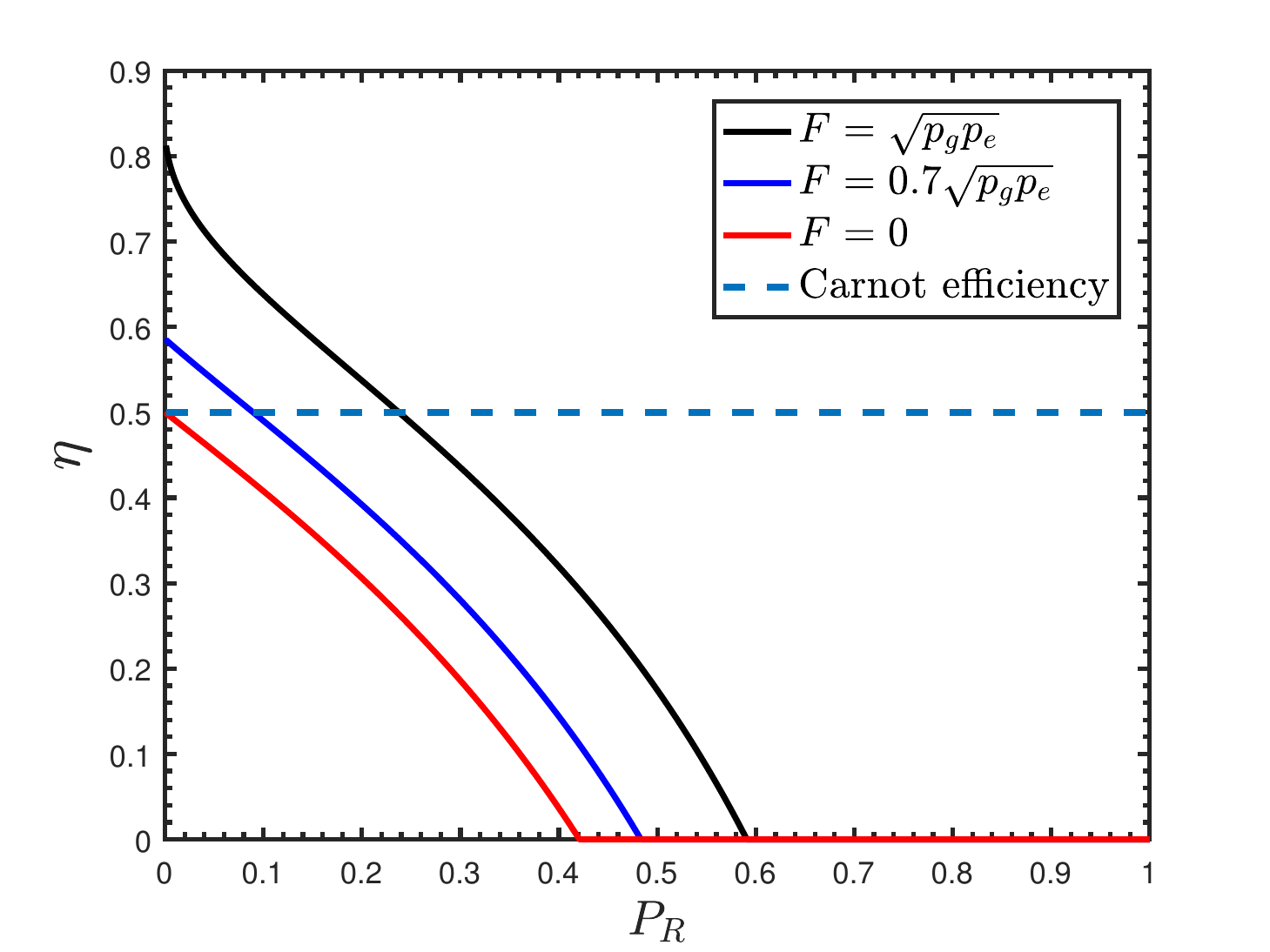}
	\caption{(color online) The efficiency $\eta$ with respect to the quantum probability $P_R$ of finding the system $S$ on the right side for different off-diagonal (coherent) element $F$. In the classical case, the efficiency must be limited by Carnot efficiency (the red line). However, with the coherence-assisted QMD, this limit may be allowed for a breakthrough (the case of pure sate with $F=\sqrt{p_gp_e}$ is depicted by black line and the blue line represents one of the mixed states with $F=0.7\sqrt{p_gp_e}$). Here we choose $T=1$, $T_D=0.5$, $\Delta=0.5$, and $F=F^{\ast}$.}
	\label{fig:eta}
\end{figure}

Now turn to the crucial efficiency, which is defined as dividing the total work done by the system to the outside by total heat absorbed from the outside, namely
\begin{equation}\label{eta}
\eta=\frac{\langle W_{\text{tot}}\rangle}{\langle Q_{\text{tot}}\rangle}=1-\frac{P_R(p_g-p_e)\Delta}{T(\Delta S_{\text{c}}+k_{\text{B}}\Delta C_r)}.
\end{equation}
Note that the coherence consumption appears in the denominator of the minuend, which implies the efficiency can be improved due to the coherence consumption of QMD. The more coherence consumed, the higher efficiency we can obtain. If there is no coherence being consumed, i.e. $\Delta C_r=0$, it will be limited by Carnot efficiency

\begin{alignat}{2}\label{eta_Carnot}
\begin{split}
\eta_{\max}&\geq \max\biggl\{\!1\!-\!\frac{P_R(p_g-p_e)\Delta}{k_{\text{B}}T\Delta S_{\text{c}}}\biggr\}\\
&=\!\!\lim_{P_R\rightarrow0}\!\textstyle\biggl\{\!1\!-\!\frac{(k_{\text{B}}T)^{-1}P_R(p_g-p_e)\Delta}{p_g\ln{p_g}+p_e\ln{p_e}-{(p_gP_L\!+p_eP_R)}\ln{(p_gP_L\!+p_eP_R)}-{(p_eP_L\!+p_gP_R)}\ln{(p_eP_L\!+p_gP_R)}}\!\biggr\}\\
&=\!\!\lim_{P_R\rightarrow0}\!\textstyle\biggl\{\!1\!-\!\frac{(k_{\text{B}}T)^{-1}P_R(p_g-p_e)\Delta}{p_g\ln{p_g}+p_e\ln{p_e}-{(p_gP_L)}\ln{(p_gP_L)}-{(p_eP_L)}\ln{(p_eP_L)}+P_R(p_g-p_e)\ln{(p_g/p_e)})}\!\biggr\}\\
&=1-\frac{\Delta}{k_{\text{B}}T\ln{(p_g/p_e)}}\\
&=1-\frac{T_D}{T},
\end{split}
\end{alignat}
where the third line follows from the Taylor expansion $(c_1+c_2x)\ln{(c_1\!+\!c_2x)}\!=\!c_1\ln{c_1}+c_2(\ln{c_1}+1)x+o(x^2)$ with $x\rightarrow0$ (here $c_1$ and $c_2$ are constants), and the last line follows from the definition of $T_D:=\Delta/[k_{\text{B}}\ln{(p_g/p_e)}]$ due to the initial probability distribution $p_g=1/(1+e^{-\beta_D\Delta})$ and $p_e=1-p_g$. This result speaks for itself, i.e., one may go beyond the classical Carnot efficiency $\eta_{\text{Carnot}}=1-T_D/T$ by consuming a certain amount of quantum coherence, see Fig.~\ref{fig:eta}. But it is worth stressing that the quantum second law always holds though the classical Carnot bound can be violated, because the bound is defined by the diagonal elements. If we define the Carnot efficiency by considering both the diagonal and off-dagonal elements of QMD, then the total efficiency of the quantum thermodynamic cycle will be limited by this so-called quantum Carnot efficiency.

In particular, there exists a critical probability $P_R^{\text{cri}}$, below which the efficiency goes beyond the classical Carnot efficiency. According to Eq.~\eqref{eta}, $P_R^{\text{cri}}$ satisfies $T_D(\Delta S_{\text{c}}+k_{\text{B}}\Delta C_r)=P_R^{\text{cri}}(p_g-p_e)\Delta$, which means QMD's absorption of heat maintains balance with the energy flow of the total composite system. Furthermore, when $P_R$ is larger than a certain value $P_R^0$, the efficiency is zero. However, with the increase of coherence, the value of $P_R^0$ will become higher. This quirk is completely the quantum effect as it depends on how much resource of quantum coherence we expend, in the light of the quantitative resource theory of coherence.
%%%%%%%%%%%%%%%%%%%%%%%%%%%%%%%%%%%%%%%%%%%%%%%%%5

\section{Heat Engine Driven by Quantum Coherence and General Coherence-Modified Second Law}
\begin{figure}[ht]
	\centering
	\includegraphics[width=4.5in]{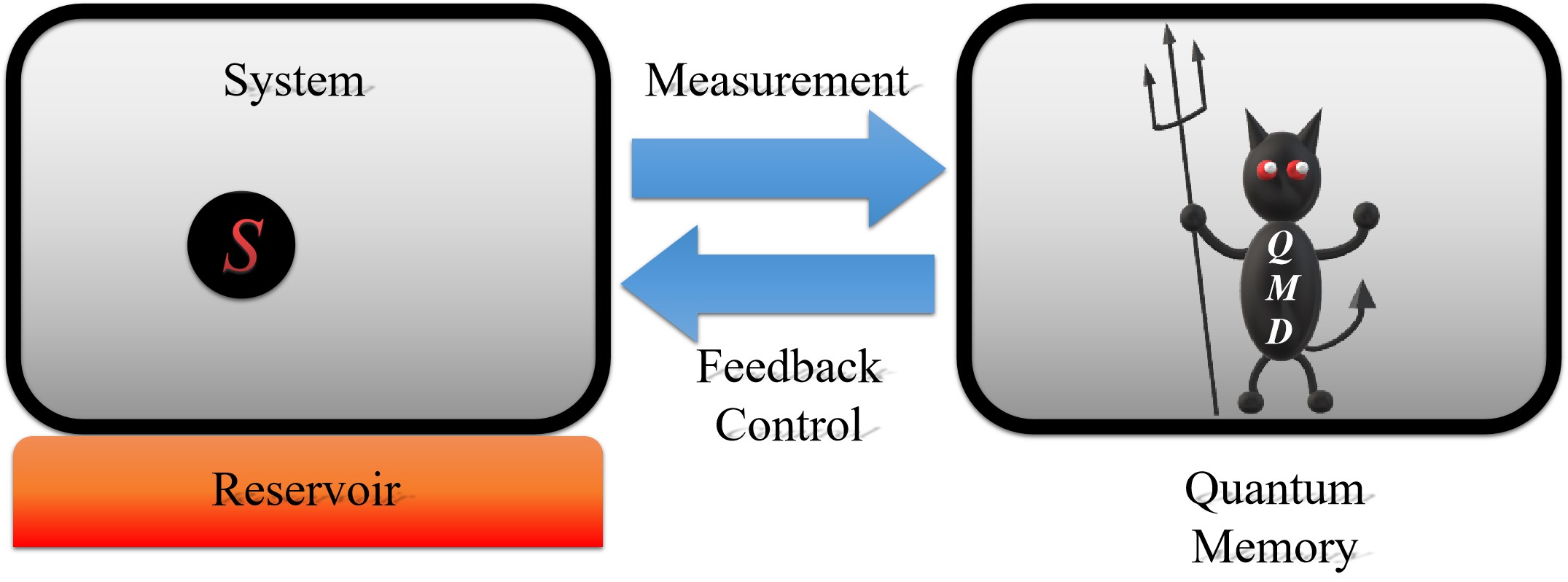}
	\caption{(color online) The schematic diagram of the information heat engine driven by quantum coherence.}
	\label{fig:IHE}
\end{figure}

Additionally, we discuss an information heat engine (IHE) driven by quantum coherence, which is inspired by Ref.~\cite{PRL-Park2013,PRA-Fan2017}. The IHE is constituted by a system $S$ and a reservoir $R$, which is controlled by a demon consisting of two memories $A$ and $B$. One can extract work from this IHE by
using quantum mutual information, or split into the classical correlation and quantum discord,
between these two memories~\cite{PRL-Park2013}. Then the IHE was generalized to multi-reservoirs case by Ren \textit{et al}. who still discussed the work extractable in the same context of quantum discord~\cite{PRA-Fan2017}.  Just according to the theme of the present work, a natural question is that whether work can be extracted from the inherent property of one quantum system---coherence, rather than the quantum correlation like discord between two or more systems. We replace the two memories with one memory $M$ in which quantum coherence exists but without any correlations, and investigate the relation between quantum coherence and work extractable from the engine. The schematic diagram of this coherent information heat engine is briefly shown in Fig.~\ref{fig:IHE}.

Stage (\romannumeral1): Initial state.---
Initially, the demon consisting of one memory $M$ is off-line and the system $S$ contacts with the reservoir $R$ at temperature $T=(k_{\text{B}}\beta)^{-1}$ in thermodynamic equilibrium, i.e. the total density matrix of the initial state is expressed as
\begin{equation}\label{rhoi}
\rho^{(i)}=\rho^{(i)}_{M}\otimes\rho^{(i)}_{SR}=\rho^{(i)}_{M}\otimes\frac{e^{-\beta H^{(i)}_S}}{Z^{(i)}_S}\otimes\frac{e^{-\beta H_R}}{Z_R},
\end{equation}
where $Z^{(i)}_S=\Tr[e^{-\beta H^{(i)}_S}]$ and $Z_R=\Tr[e^{-\beta H_R}]$ are the initial partition functions of the system $S$ and the reservoir $R$, respectively. In general, there is no restriction on $\rho^{(i)}_{M}$.

Stage (\romannumeral2): Unitary evolution.---
The system $S$ begins to interact with $R$ (a unitary evolution), and $M$ is still off-line. Then the density matrix is given by
\begin{equation}\label{rho1}
\rho^{(1)}=U^{(1)}\rho^{(i)}{U^{(1)}}^{\dagger},
\end{equation}
with $U^{(1)}=I_M\otimes U^{(1)}_{SR}$. In this stage, the memory still doesn't participate in the process of IHE.

Stage (\romannumeral3): POVM.---
This stage is where the demon consisting of $M$ started working. The measurement is performed by $M$ with POVMs (positive operator valued measures)~\cite{book-Nielsen2000,book-Sagawa2013}. Explicitly, the measurement process is implemented by performing a unitary transformation $U^{(2)}$ on the whole system followed by a projection measurement $\{\Pi_M^k: \ket{k}_M\bra{k}\}$ (a rank-1 projector) only on $M$, namley
\begin{equation}\label{rho2}
\rho^{(2)}=\sum_k\Pi_M^kU^{(2)}\rho^{(1)}{U^{(2)}}^{\dagger}\Pi_M^k=\sum_kp_k\ket{k}_M\bra{k}\otimes\rho_{SR}^{(2)k},
\end{equation}
where $p_k=\Tr[\Pi_M^kU^{(2)}\rho^{(1)}{U^{(2)}}^{\dagger}\Pi_M^k]$ is the measurement outcome registered by the memory and the postmeasurement state of $SR$ is $\rho_{SR}^{(2)k}=\Tr_A[\Pi_M^kU^{(2)}\rho^{(1)}{U^{(2)}}^{\dagger}\Pi_M^k/p_k]$.

Stage (\romannumeral4): Feedback control.---
After measurement, the demon will control $SR$ according to the outcome $p_k$. Mathematically, this feedback control is performed by a unitary operator
\begin{equation}\label{U3}
U^{(3)}=\sum_k\ket{k}_M\bra{k}\otimes U_{SR}^k,
\end{equation}
then the final state becomes
\begin{equation}\label{rhof}
\rho^{(f)}=U^{(3)}\rho^{(2)}{U^{(3)}}^{\dagger}.
\end{equation}
Note that the final state $\rho^{(f)}$ is not necessarily the canonical distribution, i.e. $\rho_{SR}^{(f)\text{can}}=\exp{(-\beta H_S^{(f)})}/Z_S^{(f)}\otimes\exp{(-\beta H_R})/Z_R$.

Since the POVMs increases the entropy while the unitary evolution ahead of the measurement keep it invariant, i.e., $S[\rho^{(2)}]\geq S[\rho^{(1)}]=S[U^{(1)}\rho^{(i)}{U^{(1)}}^{\dagger}]=S[\rho^{(i)}]$, one can obtain
\begin{equation}\label{ineq:S2>=Si}
S[\rho_{SR}^{(i)}]+S[\rho_{M}^{(i)}]\leq S[\rho_{M}^{(2)}]+\sum_kp_kS[\rho_{SR}^{(2)k}]\leq S[\rho_{M}^{(2)}]+S[\rho_{SR}^{(2)}],
\end{equation}
where the second inequality follows from the subadditivity of von Neumann entropy~\cite{book-Nielsen2000}.

Considering the final state of $SR$, i.e., $\rho_{SR}^{(f)}=\Tr_M[\rho^{(f)}]=\sum_kU_{SR}^k\rho_{SR}^{(2)k}U_{SR}^{k\dagger}$, we can obtain the following inequality by virtue of the concavity of the von Neumann entropy~\cite{book-Nielsen2000}
\begin{equation}\label{S(rhoSRf)}
S[\rho_{SR}^{(f)}]\geq\sum_kp_kS[\rho_{SR}^{(2)k}].
\end{equation}
Likewise, we have $\rho_M^{(f)}=\Tr_{SR}[\rho^{(f)}]=\sum_k\ket{k}_M\bra{k}=\Tr_{SR}[\rho^{(2)}]=\rho_M^{(2)}$, thus
\begin{equation}\label{S(rhoMf)}
S[\rho_{M}^{(f)}]=S[\rho_{M}^{(2)}].
\end{equation}
By substituting Eqs.~\eqref{S(rhoSRf)} and~\eqref{S(rhoMf)} into Eq.~\eqref{ineq:S2>=Si}, one obtains
\begin{equation}\label{ineq:entropy increase1}
S[\rho_{SR}^{(i)}]-S[\rho_{SR}^{(f)}]\leq S[\rho_{M}^{(f)}]-S[\rho_{M}^{(i)}],
\end{equation}
which implies the entropy decrease of the heat engine and the reservoir cannot exceed the entropy increase of the memory, or should say, the whole system satisfies a general principle of entropy increase, i.e., $\Delta S_{SR}+\Delta S\geq0$, where $\Delta S_{SR}=S[\rho_{SR}^{(f)}]-S[\rho_{SR}^{(i)}]$ and $\Delta S=S[\rho_{M}^{(f)}]-S[\rho_{M}^{(i)}]$. According to the quantum version of Klein's inequality, i.e. $\Tr[\rho_{SR}^{(f)}\ln\rho_{SR}^{(f)\text{can}}]\leq\Tr[\rho_{SR}^{(f)}\ln\rho_{SR}^{(f)}]$~\cite{book-Sagawa2013}, Eq.~\eqref{ineq:entropy increase1} becomes
\begin{equation}\label{ineq:entropy increase2}
S[\rho_{SR}^{(i)}]+\Tr[\rho_{SR}^{(f)}\ln\rho_{SR}^{(f)\text{can}}]\leq \Delta S.
\end{equation}

To investigate the work extractable from the heat engine, we turn the inequality above into the following form by using the canonical distributions $\rho_{SR}^{(i)}=\frac{e^{-\beta H^{(i)}_S}}{Z^{(i)}_S}\otimes\frac{e^{-\beta H_R}}{Z_R}$ and $\rho_{SR}^{(f)\text{can}}=\frac{e^{-\beta H^{(f)}_S}}{Z^{(f)}_S}\otimes\frac{e^{-\beta H_R}}{Z_R}$:
\begin{equation}\label{ineq:entropy increase3}
\langle E_S^{(i)}\rangle-\langle E_S^{(f)}\rangle+\langle E_R^{(i)}\rangle-\langle E_R^{(f)}\rangle\leq F_S^{(i)}-F_S^{(f)}+k_{\text{B}}T\Delta S
\end{equation}
\\
with $\langle E_S^{(i)}\rangle=\Tr[\rho_S^{(i)}H_S^{(i)}]$, $\langle E_S^{(f)}\rangle=\Tr[\rho_S^{(f)}H_S^{(f)}]$, $\langle E_R^{(i)}\rangle=\Tr[\rho_R^{(i)}H_R]$, $\langle E_R^{(f)}\rangle=\Tr[\rho_R^{(f)}H_R]$, $F_S^{(i)}=-k_{\text{B}}T\ln{Z_S^{(i)}}$, and $F_S^{(f)}=-k_{\text{B}}T\ln{Z_S^{(f)}}$. Here, the definition of the work extractable is $\langle W_{\text{ext}}\rangle:=-\langle \Delta E_S\rangle+\langle Q_S\rangle$, where $\langle \Delta E_S\rangle=\langle E_S^{(f)}\rangle-\langle E_S^{(i)}\rangle$ is the change of the internal energy of $S$ and $\langle Q_S\rangle=\langle E_R^{(f)}\rangle-\langle E_R^{(i)}\rangle$ is the heat exchange between $S$ and $R$, and thus we obtain
\begin{equation}\label{eq:W_ext}
\langle W_{\text{ext}}\rangle\leq -\Delta F_S+T\Delta S_{\text{c}}+k_{\text{B}}T\Delta C_{r},
\end{equation}
in which $\Delta F_S=F_S^{(f)}-F_S^{(i)}$ is the difference of free energy of the system, $\Delta S_{\text{c}}=k_{\text{B}}\Delta S_{\text{diag}}=k_{\text{B}}S[\rho_{M\text{diag}}^{(f)}]-k_{\text{B}}S[\rho_{M\text{diag}}^{(i)}]$ is the classical entropy change, $\Delta C_{r}=C_{r}[\rho_M^{(i)}]-C_{r}[\rho_M^{(f)}]$ is the coherence consumption of the memory. From Eq.~\eqref{eq:W_ext}, one can notice that if the contribution of $\Delta F_S$ is ignored, the bound of the work extractable is given by the total entropy change of the memory, which is split into the change of incoherent part and coherence consumption. Even if no classical entropy changes, i.e. $\Delta S_{\text{c}}=0$, we still find the maximum work extractable is given by coherence consumption, namely
\begin{equation}\label{eq:Coherence2}
\langle W_{\text{ext}}\rangle\leq k_{\text{B}}T\Delta C_{r},
\end{equation}
which is also in support of the resource-driven viewpoint, specifically, one can extract work from quantum coherence, a potential quantum physical resource, to drive a heat engine.
%%%%%%%%%%%%%%%%%%%%%%%%%%%%%%%%%%%%%%%%%%%%5

\section{Extending to More General Thermodynamics involving Quantum Coherence}
We consider a general quantum system described as $H\ket{\psi_n}=E_n\ket{\psi_n}$, where $H$, $E_n$ and $\ket{\psi_n}$ are the Hamiltonian, $n$-th eigenenergy and eigenstate, respectively. Here, the system is not necessarily to be a single system. It might be composed of multiple subsystems. In the energy representation, a general density matrix of this system can be given by
\begin{equation}\label{rho}
\rho=\sum_nP_n\ket{\psi_n}\bra{\psi_n}+\sum_{n\neq m}P_{nm}\ket{\psi_n}\bra{\psi_m},
\end{equation}
then the internal energy of the system can be expressed as
\begin{equation}\label{U}
\langle E\rangle=\Tr[\rho H]=\sum_nP_nE_n
\end{equation}
where $P_n$ is the probability distribution of $n$-th energy eigenstate. In equilibrium $P_n$ obeys the canonical distribution. Note that although the density matrix contains off-diagonal part, the internal energy of the system is only connected with the diagonal part due to the energy representation.

%And if be not added that we will always choose the energy representation (basis) in this paper.
%We emphasize that this conclusion applies only to the case of energy representation, the situation will absolutely make difference if the representation has been transformed.

From the derivative of $\langle E\rangle$, one obtains $\D\langle E\rangle=\sum_n(E_n\D P_n+P_n\D E_n)$. Here, the previous viewpoint is analogizing it to the classical thermodynamic first law, i.e., $\D W_{\text{c}}=-\D E_{\text{c}}+\D Q_{\text{c}}=-\D E_{\text{c}}+T\D S_{\text{c}}$, where $E_{\text{c}}$, $W_{\text{c}}$, $Q_{\text{c}}$ and $S_{\text{c}}$ are the classical internal energy, work, heat and entropy, respectively. Then, the quantum work can be identified as $\D\langle W\rangle=-\sum_nP_n\D E_n$~\cite{PRL-Kieu2004,PRE-Esposito2006,PRL-Kim2011}, and quantum heat is $\D\langle Q\rangle=\sum_nE_n\D P_n$ associated with $T\D S_{\text{c}}$ since the classical entropy $S_{\text{c}}$ is defined as $S_{\text{c}}=-k_{\text{B}}\sum_nP_n\ln{P_n}$.

Nevertheless, one has already noticed that the classical entropy is actually the von Neumann entropy of diagonal density matrix in energy representation, namely
\begin{equation}\label{Sc}
S_{\text{c}}=k_{\text{B}}S(\rho_{\text{diag}}),
\end{equation}
where $k_{\text{B}}$ is the Boltzmanns constant and $\rho_{\text{diag}}=\sum_n\bra{\psi_n}\rho\ket{\psi_n}\ket{\psi_n}\bra{\psi_n}$ denotes the diagonal part of $\rho$. Therefore, with the analogy above, this quantum system must have lost part of the information concerning its non-equilibrium (off-diagonal) state. To describe the whole quantum state, the entropy must be the whole $k_{\text{B}}S(\rho)$ instead of $k_{\text{B}}S(\rho_{\text{diag}})$ which contains only the information of equilibrium (diagonal). Hence, combined with Eq.~\eqref{eq:Cre}, it is natural to define the quantum heat as
\begin{equation}\label{quanQ}
\langle Q\rangle=\int\limits_{\rho\rightarrow\sigma}\sum_nE_n\D P_n+k_{\text{B}}T\Delta C_r,
\end{equation}
where $\Delta C_r=C_r(\rho)-C_r(\sigma)$ is the coherence consumption of the system from the initial state $\rho$ to the final state $\sigma$. Then the first term of the right-hand side of Eq.~\eqref{quanQ} only determined by the diagonal (incoherent) part $P_n$ can be considered as the incoherent heat, i.e., $\langle Q_{\text{incoh}}\rangle=\int_{\rho\rightarrow\sigma}\sum_nE_n\D P_n=TS_{\text{c}}=k_{\text{B}}TS(\rho_{\text{diag}})$, and the second term can be understood as the coherent heat, i.e., $\langle Q_{\text{coh}}\rangle=k_{\text{B}}T\Delta C_r$. Thus,
\begin{equation}\label{quanQ=incoh+coh}
\langle Q\rangle=\langle Q_{\text{incoh}}\rangle+\langle Q_{\text{coh}}\rangle,
\end{equation}
which gives the physical origin of quantum heat: the change of diagonal (incoherent) distribution and the coherence consumption.

By virtue of the definition of quantum heat in Eq.~\eqref{quanQ}, we can define the quantum work as
\begin{equation}\label{quanW}
\langle W\rangle=-\int\limits_{\rho\rightarrow\sigma}\sum_nP_n\D E_n+k_{\text{B}}T\Delta C_r,
\end{equation}
since the first law $\langle \Delta E\rangle=-\langle W\rangle+\langle Q\rangle$. Likewise, the quantum work can be written as
\begin{equation}\label{quanW=incoh+coh}
\langle W\rangle=\langle W_{\text{incoh}}\rangle+\langle W_{\text{coh}}\rangle,
\end{equation}
with the incoherent work $\langle W_{\text{incoh}}\rangle=-\int_{\rho\rightarrow\sigma}\sum_nP_n\D E_n$ and the coherent work $\langle W_{\text{coh}}\rangle=k_{\text{B}}T\Delta C_r$. So the quantum work can be viewed as the contribution of both the change of energy level under the invariable diagonal (incoherent) distribution and the coherence consumption.

In this way, the total internal energy don't concern the off-diagonal part. However, because the total work can be enhanced due to the coherent superposition, the efficiency of a quantum heat engine thus can be improved in the thermodynamic process involving quantum coherence.
%%%%%%%%%%%%%%%%%%%%%%%%%%%%%%%%%%%%%%%%%%%%%%%

\section{Conclusions}
In summary, we propose and study quantum thermodynamics by utilizing the resource theory of coherence. Two kinds of quantum heat engine assisted by a coherent QMD are discussed in details, which are based on QSE and IHE, respectively. The quantum thermodynamic cycle based on QSE is divided into five stages: initial state, insertion, measurement, expansion and removal, which are all described by the evolution of quantum ensembles. We explicitly calculated the total quantum work, heat and
the corresponding efficiency of this thermodynamic cycle. From a resource-driven viewpoint, the efficiency can be enhanced due to the coherence consumption of QMD, which is one of our main results.

In addition, we discuss an universal engine driven by quantum coherence based on IHE, which is also to achieve the whole measurement and feedback control through QMD. The maximum work extractable is given by the consumption of quantum coherence, leading to a coherence-modified second law. Consequently, one can extract work from quantum coherence to drive this heat engine even without any classical resources.

Finally, the subtle connection between coherence and fundamental thermodynamic notions is extended to a more general quantum thermodynamics. The quantum work and heat can be naturally redefined as a sum of incoherent and coherent parts by considering the first and second thermodynamic laws.
Our results are enlighten for the quantum information processes in thermodynamic systems where the coherent superposition of states cannot be ignored.

\section*{Acknowledgments}
Yun-Hao Shi thanks Z. C. Tu and Shi-Ping Zeng for their valuable discussions. This work was supported by National Key Research and Development Program of China (Grant Nos. 2016YFA0302104, 2016YFA0300600), NSFC (Grants Nos. 11774406, 11847306 and 11705146), Strategic Priority Research Program of Chinese Academy of Sciences (Grant No. XDB28000000), the Key Innovative Research Team of Quantum Many-body theory and Quantum Control in Shaanxi Province (Grant No. 2017KCT-12) and the Major Basic Research Program of Natural Science of Shaanxi Province (Grant No. 2017ZDJC-32). Hu was supported by NSFC (Grant No. 11675129), New Star Project of Science and Technology of Shaanxi Province (Grant No. 2016KJXX-27) and New Star Team of XUPT.

\end{document}